\documentclass[twocolumn,showpacs,preprintnumbers,amsmath,amssymb,aps,prl]{revtex4}
\usepackage{graphicx}
\begin{document}

\title{Hysteresis and Return Point Memory in Artificial Spin Ice Systems}   
\author{A. Lib{\' a}l$^{1}$, C. Reichhardt$^{2}$, and    
 C. J. Olson Reichhardt$^{2}$} 
\affiliation{
$^1$Faculty of Mathematics and Computer Science,
Babes-Bolyai University, RO-400591 Cluj-Napoca, Romania\\
$^2$ Theoretical Division,
Los Alamos National Laboratory, Los Alamos, New Mexico 87545, USA } 

\date{\today}
\begin{abstract}
We investigate hysteresis loops and return point memory 
for artificial square and kagome spin ice systems by
cycling an applied bias force and comparing 
microscopic effective spin configurations throughout the 
hysteresis cycle. 
Return point memory loss is caused by motion of individual defects
in kagome ice or of grain boundaries in square ice.
In successive cycles, return point memory is recovered
rapidly in kagome ice.  Memory is 
recovered more gradually in square ice due to the
extended nature of the grain boundaries.
Increasing the amount of quenched disorder increases the defect density but
also enhances the
return point memory since the defects become trapped more easily.
\end{abstract}
\maketitle

\vskip2pc
Frustration effects arise 
in many
condensed and soft matter systems   
when geometric constraints prevent
collections of interacting elements such as spins or charged particles 
from simultaneously minimizing all pairwise interaction energies.  
One of the best known 
frustrated systems
are the spin ices \cite{Anderson,SpinIce}, named
for their similarity to the frustrated proton ordering 
in water ice \cite{Pauling}. 
Spin ices have been realized in both two and three dimensions and  
exhibit interesting 
excitations
such as effective magnetic monopoles \cite{Anderson,Monopole}. 
More recently, 
artificial spin ices were created with
arrays of nanomagnets 
\cite{ShifferMollerNisoliLi,GB,EM,Zabel,OT,Cumings,Tanaka,Mo,S,Pkohli},
colloidal particles \cite{Libal,Colloid}, 
or vortices in nanostructured superconductors \cite{Vortex}.   
In artificial ices,
direct visualization of the microscopic 
effective spin configurations is possible, and system parameters
such as interaction strength, doping, or the amount of quenched disorder
can be controlled. 
Under a varying external field, changes in the microscopic configurations
can be imaged
and used to construct
hysteresis loops \cite{OT,Zabel,EM,Pkohli}, as shown
for kagome ice 
where the motion, creation, and annihilation of topological defects  
along the hysteresis cycle were demonstrated \cite{EM}.   
Memory effects are generally associated with hysteresis, and in
return point memory (RPM),
the system returns to the same 
{\it microscopic} configuration after completing a hysteresis loop
\cite{Dahmen,P,MP,Narayan,Zimayi}.
Recently developed techniques show that in real magnetic materials,
RPM occurs in strongly disordered samples
and is absent for weak disorder \cite{P,MP,Zimayi}.     
Certain classes of $T=0$ 
disordered spin systems, 
such as the random field Ising model \cite{Dahmen}, 
exhibit perfect RPM, while 
other systems require many 
loops to organize into a state with RPM \cite{Narayan,MP}. 

Artificial spin ices are an ideal system for studying
RPM since they exhibit 
hysteresis and the microscopic states can be visualized directly.
The type of topological defect that forms and its mobility 
varies in different
ice systems,
ranging from mobile monopoles \cite{OT,EM} in kagome ice
to less mobile grain boundaries \cite{Vortex,GB} in square
ice, 
and this could modify the RPM behavior.
To 
quantify this, we perform numerical simulations 
of hysteresis in artificial square and kagome spin ices constructed
from colloids in double-well traps 
with varied amounts of 
quenched disorder.
Our model was previously shown to capture the behavior 
of square and kagome ices \cite{Libal,Vortex}, 
and the number
and type of topological defects present 
can be controlled by 
changing the amount of quenched disorder \cite{Vortex}. 
We using molecular dynamics simulations
to avoid      
the spurious loss of RPM that occurs 
under Monte Carlo spin flip methods \cite{Zimayi}. 
Our work implies that RPM phenomena can be 
studied in general artificial spin systems
and not just artificial spin ice systems, 
which would provide a new method for exploring
microscopic memory effects in condensed matter systems.  

{\it Simulation--}
We simulate an artificial spin ice of $N$
charged colloidal particles trapped in an array 
of elongated
double-well 
pinning sites that have two states 
determined by
which 
well 
is occupied by the colloid.
The dynamics of colloid $i$ is governed by the
overdamped equation of motion
$\eta(d{\bf R}_i/dt)={\bf F}_i^{cc}+{\bf F}_i^s+{\bf F}_{\rm ext}$,
where the damping constant $\eta=1$.
The colloid-colloid interaction force has a Yukawa or
screened Coulomb form,
${\bf F}_{i}^{cc} = -F_0q^2\sum^{N}_{i\neq j}\nabla_i V(R_{ij})$,
with
$V(R_{ij}) = (1/R_{ij})\exp(-\kappa R_{ij}){\bf {\hat r}}_{ij}$.
Here $R_{ij}=|{\bf R}_{i} - {\bf R}_{j}|$,
${\bf {\hat R}}_{ij}=({\bf R}_{i}-{\bf R}_{j})/R_{ij}$,
${\bf R}_{i(j)}$ is the position of particle $i$($j$),
$F_0=Z^{*2}/(4\pi\epsilon\epsilon_0)$,
$Z^*$ is the unit of charge, $\epsilon$ is the
solvent dielectric constant,
$q$ is the dimensionless colloid charge,
$1/\kappa=4a_0$ is the screening length,
and $a_0$ is the unit of distance that is typically of order a micron.
We neglect hydrodynamic interactions between colloids since we work in
the low volume fraction limit and the colloids remain confined in the pins.
The pinning force ${\bf F}_s$ arises from $N_p$ elongated traps of
length $l=1.333a_0$, width $d_p=0.4a_0$, and depth $f_p=100F_0$. 
The pin ends are parabolic confining potentials 
with radius $r_p=0.2a_0$.
A cylindrical force restricts motion 
in the direction perpendicular to the long axis of the pinning site, and
a barrier in the center of the pinning site is produced by a
repulsive parabolic force of height 
$f_r$
that creates two energy minima on either end of
the pin \cite{Libal}.
For square ice the pins are arranged 
with $v=4$ traps meeting at each vertex, as shown 
in Fig. 1, while for kagome ice, $v=3$ traps meet at
each vertex \cite{Vortex}.
The distance between adjacent vertices is $a=2a_0$ and there are $N_v$ vertices.
Our square ice has $35 \times 35$ vertices
($N_v=1225$) and $N_p=2450$ elongated pins, 
while our kagome ice has
$40 \times 40$ vertices ($N_v=1600$) 
and $N_p=2400$ elongated pins. 
Disorder is added to the system by increasing or decreasing 
$f_r$ in individual pinning sites according to a normal distribution
with mean $f_b=1.0F_0$ and standard deviation $\sigma$.
This is analogous to varied island coercive fields
in the nanomagnetic system.
We initialize the system by placing a colloid in one randomly selected
end of each pinning site so that $N=N_p$.
To construct a hysteresis loop we apply an  
external force ${\bf F}_{\rm ext}$ uniformly to the sample, which for charged
colloidal particles could be achieved using an external electric field.
In the kagome ice ${\bf F}_{\rm ext}=F_{\rm ext}{\bf \hat x}$ 
while in the square ice 
${\bf F}_{\rm ext}=F_{\rm ext}\sqrt{2}({\bf \hat x}+{\bf \hat y})/2$, 
as shown in Fig.~1.
We sweep $F_{\rm ext}$ from zero to a positive maximum value $F_{\rm max}$, then
back down through zero to a negative maximum value $-F_{\rm max}$, and finally
back up to zero to create one loop.

\begin{figure}
\includegraphics[angle=-90,width=3.5in]{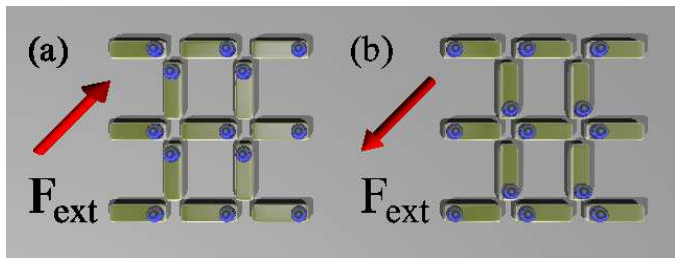}
\caption{ 
Schematic of a portion of the square ice sample.  A charged colloid
(dots) can sit in either end of each trap (rectangles).
(a) A biased state for $F_{\rm ext}$ applied at $\theta=45^{\circ}$ from
the $x$ axis,
where all the vertices are in the $(0011)$
configuration. 
(b) The negative biased state
after the drive is reversed 
contains only $(1100)$ vertices. 
By cycling
$F_{\rm ext}$ between these two extremes we construct
an effective magnetization curve.
}
\label{fig:schem}
\end{figure}

The vertices formed by the meeting points of the 
pins
are categorized by $n_{in}=\sum_{i=1}^vc_i$, where $c_i=1$ if a
colloid is sitting in the end of the pin closest to the vertex and $c_i=0$
if the colloid is sitting in the end of the pin furthest from the vertex
\cite{Libal,Vortex}. 
Each possible vertex type can be written as $(c_1c_2c_3c_4)$ for the square ice
and as $(c_1c_2c_3)$ for the kagome ice, starting with the pinning
site lying along the positive $x$ axis with respect to the vertex and proceeding
counterclockwise around the vertex.
The $n_{in}=2$ vertices in square ice do not all have the same energy.
We denote
the ground state vertices (1010) and (0101) as type $2gs$, and
the higher energy biased vertices (1100), (0110), (0011), and (1001)
as type $2b$.

\begin{figure}
\includegraphics[width=3.5in]{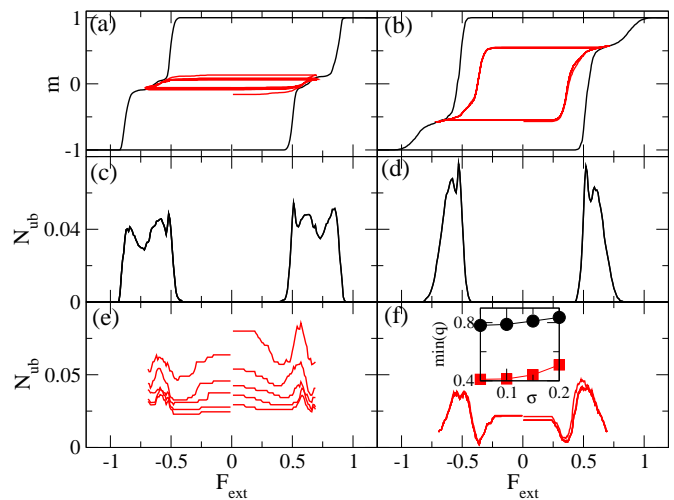}
\caption{
(a,c,e) Square ice sample with $\sigma=0.1$. (b,d,f) Kagome ice 
sample with $\sigma=0.1$.
(a,b) The reduced magnetization $m$ vs $F_{\rm ext}$.
Saturation occurs at $m= \pm 1.0$ when all the vertices are in 
biased states. 
Dark line: Saturated loop with
$F_{\rm max}=2.0$.
Light lines: Consecutive loops with $F_{\rm max}=0.7$, below saturation. 
The virgin curves are not shown.
(c,d) Fraction of unbiased vertices $N_{ub}$ 
vs $F_{\rm ext}$ for the saturated loop with $F_{\rm max}=2.0$.
(e,f) $N_{ub}$ vs $F_{\rm ext}$ for repeated unsaturated loops
with $F_{\rm max}=0.7$, with cycle number $n$ increasing from top to
bottom.
For clarity, we omit the horizontal lines connecting 
$F_{\rm ext}=\pm F_{\rm max}$ to $F_{\rm ext}=0$.
There is a much greater decrease in $N_{ub}$ for the square ice than for
the kagome ice.
Inset of (f): min$(q)$, the effective spin overlap in the $n=2$ cycle vs
$\sigma$ for (circles) kagome and (squares) square ice.  Samples
with stronger disorder have higher $q$ values.
}
\label{fig:mag}
\end{figure}

We first show that our model captures the hysteretic behavior observed
in artificial ice systems \cite{OT,Zabel,EM,P}. 
In the absence of quenched disorder or drive, we find 
ice-rule obeying states that are ordered ground states
in the square ice
\cite{Libal} and 
disordered 
in the kagome ice.
When we add quenched disorder with $\sigma>0$, due to its lack of 
extensive degeneracy the square ice forms grain
boundaries composed of non-ice rule obeying vertices as shown
in simulation \cite{Vortex} and experiment \cite{GB}, 
while in kagome ice isolated non-ice rule defects appear \cite{Vortex,EM}. 
Under the biasing drive ${\bf F}_{ext}$ shown schematically in Fig.~1(a),
all the square ice vertices adopt the $(0011)$ configuration. 
For kagome ice biased along the $x$ direction, 
all the vertices enter the $(011)$ state.
We define the reduced magnetization $m$ 
as the fraction of vertices aligned with
the saturation direction, $M=N_v^{-1}\sum_{i=1}^{N_v}s_i$, where $s_i=1$ for
$(0011)$ vertices in square ice or $(011)$ vertices in kagome ice,
$s_i=-1$ for $(1100)$ vertices in square ice or $(110)$ vertices in kagome
ice, and $s_i=0$ for all other vertices.
In Fig.~2(a,b) we plot the hysteresis loops for square 
and kagome ice samples with 
$\sigma=0.1$. 
The thick curve is obtained with $F_{max}=2.0$, beyond the saturation
level where $m=\pm 1$.
We plot the fraction of unbiased vertices,
$N_{ub}=N_v^{-1}\sum_{i=1}^{N_v}\delta(s_i)$, vs $F_{\rm ext}$ with
$F_{\rm max}=2.0$ 
in Fig.~2(c,d). 
The completely ordered biased states are only destroyed
for $0.4 < |F_{ext}| < 1.0$,
close to the coercive fields 
at which the effective spin direction flips.
The shape of the hysteresis loop and the peaks in the 
non-biased defect density in Figs.~2(b,d) are in excellent agreement with 
the digitally constructed hysteresis loops produced in experiments on
nanomagnetic kagome ice samples \cite{OT,Zabel,EM}. 
The thin lines in Figs.~2(a,b) show consecutive hysteresis loops 
obtained below saturation with $F_{\rm max}=0.7$, near
the middle of the range of $F_{\rm ext}$ in which 
the largest number of
defects appear.
In Fig.~2(e,f) we plot $N_{ub}$ versus $F_{\rm ext}$ 
for the unsaturated hysteresis loops.
For the square ice, Fig.~2(e) shows
that $N_{ub}$ decreases with increasing $n$, where $n$ is
the number of loops performed, 
indicating that defect annihilation is occurring. 
For continued cycling beyond the number of loops shown in the figure,
the system settles into a steady state. 
In the kagome ice, Fig.~2(f) shows that
$N_{ub}$ hardly changes from one cycle
to the next, indicating that only a small number of defects annihilate. 
For the saturated case with $F_{max} = 2.0$ shown in Fig.~2(b,c),
the $N_{ub}$ curves do not evolve under repeated looping since the sample
loses all memory of the microscopic configuration near the coercive field
once saturation is reached.

\begin{figure}
\includegraphics[width=3.5in]{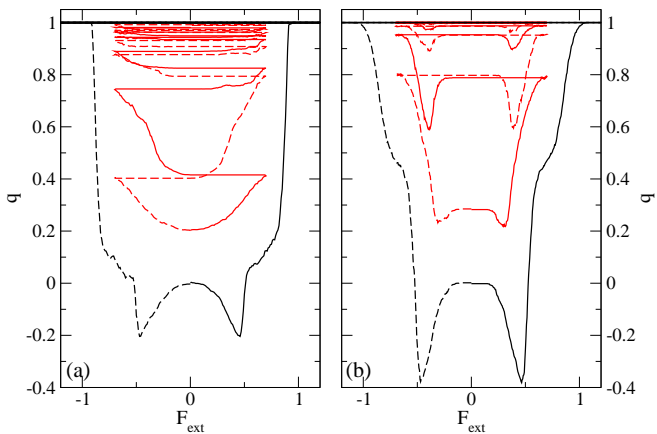}
\caption{ 
Effective spin overlap $q$ vs $F_{\rm ext}$ during consecutive
hysteresis loops for (a) a square ice sample with $\sigma=0.1$ and (b)
a kagome ice sample with $\sigma=0.1$.  Heavy line: 
Saturated loops with $F_{\rm max}=2.0$, including the
virgin curve.  Thin lines: Unsaturated loops with $F_{\rm max}=0.7$, with
$n$ increasing from bottom to top.
Solid lines: clockwise loops; dashed lines: counterclockwise loops.
In the kagome ice, $q$ approaches 1 after only a few cycles, while a much
larger number of cycles are required before $q$ approaches 1 in the square
ice.
}
\label{fig:rpm}
\end{figure}

We quantify the RPM by measuring the 
overlap $q$ in the effective spin configurations along
a hysteresis loop \cite{MP,Zimayi} at equal
values of $F_{\rm ext}$ after $n$ complete cycles.  For each trap, we define
an effective spin $S_i=1$ if the colloid is sitting in the right or top
end of the trap, and $S_i=-1$ if the colloid is sitting in
the left or bottom end of the trap.  Writing the 
value of $S_i$ after $n$ cycles as $S_i^{(n)}$, we measure
\begin{equation}
q(F_{\rm ext}) = {N}^{-1}\sum^{N}_{i=1}S_{i}^{(n-1)}(F_{\rm ext})S_{i}^{(n)}(F_{\rm ext}) .
\end{equation}   
The term in the sum is 1 if the trap was
biased in the same direction both before and after the complete cycle, and -1
if the colloid jumped to the other end of the trap.  
In Fig.~3(a,b) we plot $q$ versus $F_{\rm ext}$ for both the saturated
and unsaturated hysteresis
curves in the square and kagome ices shown in Fig.~2(a,b). 
In the case of the saturated loops, $q$ for the virgin curve 
in Fig.~3(a) shows that
since the sample was not initialized in a biased state, the initial
configuration differs significantly from the effective spin configuration
obtained one cycle later, but for $n=2$ and above, $q=1$, indicating
perfect memory.
For the unsaturated loops obtained with $F_{\rm max}=0.7$, $q$ is low during
the first cycle, but as $n$ increases $q$ gradually converges to a value
just below $q=1$.
A comparison with Fig.~2(e) indicates that
the increase in memory with increasing $n$ is correlated with
a decrease in $N_{ub}$, although for this value of $F_{\rm max}$ there are
always some defected vertices present even after the system reaches
a steady state in which the grain boundaries cease to evolve.
The kagome ice in Fig.~3(b) shows a similar behavior except
that $q$ approaches 1 after only a few cycles, leading to a much faster
establishment of RPM than in the square ice.
In Fig.~2(f) we show that the number of defected vertices remains
nearly constant in the kagome ice even under repeated cycling.
This indicates that although the kagome ice defects do not annihilate,
they are mobile during the first few cycles and then become pinned.
Our results demonstrate that for the square ice, changes in the amount of RPM are
primarily associated with the annihilation of defects, while in the 
kagome ice, RPM is suppressed by the motion of defects.

Although the number of defects $N_{ub}$ in both types of ice increases 
with increasing disorder $\sigma$, 
the amount of RPM increases with increasing disorder.
We illustrate this in the inset of Fig.~2(f) 
where we plot the value of $q$ on the $n=2$ plateau versus 
$\sigma$.
A similar effect was observed for 
real magnetic systems and in spin simulations \cite{P,MP,Zimayi}. 
In our system, $q$ increases with increasing disorder 
due to the stronger pinning of the domain walls in the square ice
or of the individual defects in the kagome ice.
In the square ice, the disorder prevents the domain walls from coarsening.
It was previously shown that as the particle-particle interaction 
strength in our system is reduced, non-ice-rule obeying vertices begin to
appear since their energetic cost decreases \cite{Libal}.
For noninteracting colloids, the sample is strongly disordered but also has
perfect RPM since the defected configurations are controlled only by the 
local disorder and are not modified by particle interactions.
Thus we expect that in the experimental nanomagnetic artificial ices,
when the coupling is reduced for increased spacing between the nanomagnets,
the system should show increased or perfect RPM.

\begin{figure}
\includegraphics[width=3.5in]{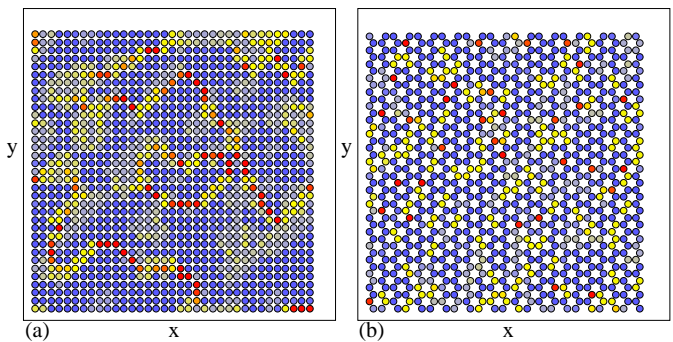}
\caption{
Vertices in (a) square ice and (b) kagome ice samples from Fig.~2 during
repeated hysteresis cycles.
Vertices are colored depending on how many cycles the vertex spends
in a defect state, ranging from dark blue for never defected sites to dark
red for permanently defected.
In (a), motion and annihilation of defects occur near grain boundaries.
In (b), individual defects move and are pinned independently without forming
grain boundaries.
}
\label{fig:diverge2}
\end{figure}

To illustrate the defect dynamics,
in Fig.~4 we plot the vertices colored according to the number of
hysteresis cycles each vertex spent as a defected site.  Red vertices
indicate locations where defects became trapped.
Figure~4(a) shows that in the square ice,
the defects organize into grain boundaries which move and coarsen under
repeated hysteresis cycles.
In Fig.~4(b), the kagome ice contains no grain boundaries but has a smaller
fraction of intermediately colored vertices
compared to the square ice since the isolated defects
become trapped after only a few cycles.
The square ice grain boundaries are less well pinned than the isolated kagome
defects since they are extended objects.
The motion of individual defects in kagome ice has already been 
imaged in experiments; it would be interesting to observe whether
these defects become localized within a few hysteresis cycles.

For most artificial ice systems, thermal effects are not relevant; 
however, thermal fluctuations can be significant in a colloidal system. 
We find that our results are robust against the addition of weak thermal
disorder, and that for $T>0$ there is only 
a slight reduction in the asymptotic value of $q$ and
a slight increase in the number of cycles required to reach a steady state.
For higher temperatures, RPM is lost even when the system is cycled to
saturation
since the thermal fluctuations cause random effective spin flips that change 
the path on each cycle.  There is also no increase 
in RPM under repeated cycles at higher temperature \cite{Zimayi}.     

In summary, we have 
studied hysteresis and return point memory effects for 
artificial square and kagome ices at the microscopic level. 
In the square ice 
for repeated unsaturated hysteresis loop cycles 
that extend to biases near the coercive field,
the RPM increases with each cycle as the grain boundaries present in the
sample coarsen and become pinned.
In kagome ice the number of defects remains nearly constant under repeated
hysteresis cycles and there is much higher RPM.
Here, individual defects hop rather than annihilating 
and are eventually pinned at sites with stronger disorder.
The grain boundaries in the square ice are more mobile than the individual
defects in the kagome ice since they are extended objects.
Our results
can be tested readily in different types of artificial ices 
and also could be studied in more general artificial
spin systems.

This work was carried out under the auspices of the 
NNSA of the 
U.S. DoE
at 
LANL
under Contract No.
DE-AC52-06NA25396.

\end{document}